\documentclass{iau307}
\usepackage{graphicx}
\usepackage{natbib}
\usepackage{url}
\usepackage{dtklogos}
\bibpunct{(}{)}{;}{a}{}{,}

\title[Rotational velocities of the O-type stars in 30\,Dor] %% give here short title %%
{Rotational velocities of single and binary O-type stars in the Tarantula Nebula}

\author[Ram\'irez-Agudelo et al.]   %% give here short author list %%
{O.H. Ram\'irez-Agudelo$^1$,
H. Sana$^2$,
A. de Koter$^{1,3}$,
S. Sim\'on-D\'{i}az$^{4,5}$,
S.E. de Mink$^{6,7}$,
F. Tramper$^{1}$,
P.L. Dufton$^{8}$,
C.J. Evans$^{9}$,
G. Gr\"afener$^{10}$,
A. Herrero$^{4,5}$,
N. Langer$^{11}$,
D.J. Lennon$^{12}$,
J. Ma\'{i}z Apell\'aniz$^{13}$,
N. Markova$^{14}$,
F. Najarro$^{15}$,     
J. Puls$^{16}$,
W.D. Taylor$^{9}$,      
\and J.S. Vink$^{10}$
 }

\affiliation{$^1$Astronomical Institute Anton Pannekoek, University of Amsterdam, The Netherlands \\ email: {\tt o.h.ramirezagudelo@uva.nl} \\[\affilskip]
$^2$ESA/Space Telescope Science Institute 3700 San Martin Drive, Baltimore, MD21218, USA
$^3$Instituut voor Sterrenkunde, Universiteit Leuven, Celestijnenlaan 200D, 3001, Leuven, Belgium
$^4$ Instituto de Astrof\'{i}sica de Canarias, C/ V\'{i}a L\'{a}ctea s/n, E-38200 La Laguna, Tenerife, Spain 
$^5$ Departamento de Astrof\'{i}sica,  Universidad de La Laguna,  Avda. Astrof\'{i}sico Francisco S\'{a}nchez s/n, E-38071 La Laguna, Tenerife, Spain
$^6$ Observatories of the Carnegie Institution for Science, 813 Santa Barbara St, Pasadena, CA 91101, USA
$^7$ Cahill Center for Astrophysics, California Institute of Technology, Pasadena, CA 91125, USA
$^8$ Astrophysics Research Centre, School of Mathematics and Physics, Queen's University of Belfast, Belfast BT7 1NN, UK 
$^9$ UK Astronomy Technology Centre, Royal Observatory Edinburgh, Blackford Hill, Edinburgh, EH9 3HJ, UK           
$^{10}$ Armagh Observatory, College Hill, Armagh, BT61 9DG, Northern Ireland, UK 
$^{11}$ Argelander-Institut f\"ur Astronomie, Universit\"at Bonn, Auf dem H\"ugel 71, 53121 Bonn, Germany 
$^{12}$ European Space Agency, European Space Astronomy Centre, Camino Bajo del Castillo s/n, Urbanizaci—n Villafranca del Castillo,
28691 Villanueva de la Ca–ada, Madrid, Spain
$^{13}$ Instituto de Astrof\'{i}sica de Andaluc\'{i}a-CSIC, Glorieta de la Astronom\'ia s/n, E-18008 Granada, Spain
$^{14}$ Institute of Astronomy with NAO, Bulgarian Academy of Science, PO Box 136, 4700 Smoljan, Bulgaria
$^{15}$ Centro de Astrobiolog\'{i}a (CSIC-INTA), Ctra. de Torrej\'on a Ajalvir km-4, E-28850 Torrej\'on de Ardoz, Madrid, Spain
$^{16}$ Universit\"atssternwarte, Scheinerstrasse 1, 81679 M\"unchen, Germany
}
\pubyear{2014}
\volume{307} 
\pagerange{}
\jname{New windows on massive stars: asteroseismology, interferometry, and spectropolarimetry}
\editors{G. Meynet, C. Georgy, J.H. Groh \& Ph. Stee, eds.}
\begin{document}

\maketitle

\begin{abstract}
Rotation is a key parameter in the evolution of massive stars, affecting their evolution, chemical yields, ionizing photon budget, and final fate.
We determined the projected rotational velocity, $v_e\sin i$, of $\sim$330 O-type objects, i.e. $\sim$210 spectroscopic single stars and $\sim$110 primaries in binary systems,
in the Tarantula nebula or 30 Doradus (30\,Dor) region. % , the starbursting main region in the Large Magellanic Cloud.  
The observations were taken using VLT/FLAMES and constitute the largest homogeneous dataset of multi-epoch 
spectroscopy of O-type stars currently available.
The most distinctive feature of the $v_e\sin i$ distributions of the presumed-single stars and primaries
in 30 Dor is a low-velocity peak at around 100\,$\rm{km s^{-1}}$.  Stellar winds are not expected to have spun-down the bulk of the
stars significantly since their arrival on the main sequence and therefore the peak in the single star sample 
is likely to represent the outcome of the formation process.
Whereas the spin distribution of presumed-single stars shows a well developed tail of stars rotating more rapidly than
300\,$\rm{km s^{-1}}$, the sample of primaries does not feature such a high-velocity tail.
%In the sample of primaries there is no significant contribution of stars rotating more rapidly than 300\,$\rm{km s^{-1}}$,
%whereas the presumed-single stars do show a well developed high-velocity tail. 
The tail of the presumed-single star distribution is attributed for the most
part -- and could potentially be completely due -- to spun-up 
binary products that appear as single stars or that have merged.  This would be consistent with the lack of such 
post-interaction products in the binary sample, that is  
% and has important implications for the evolutionary origin of the progenitors of long gamma-ray bursts.
%where is difficult to detect the binary system or there is no longer a binary due to merging of the two components.
%The binary sample is 
expected to be dominated by pre-interaction systems. 
The peak in this distribution is broader and is shifted toward somewhat higher spin rates
compared to the distribution of presumed-single stars.  Systems displaying large radial velocity variations, typical for short period systems, appear mostly responsible for these differences.  
%The spin rates of the primaries and secondaries of these large radial velocity systems tend to be similar, suggesting tidal locking.
%Tidal interaction may explain the locking of spin rates of primaries.
%Tidal interaction may explain the locking of the spin rates of primaries. 
%The lack of binaries displaying large spin rates is also 
%consistent with the idea that most such systems are post-interaction binaries.
%in which the secondary Ð dominating the light Ð is spun up dramatically but displays small radial velocity variations.

%, line: profiles

\end{abstract}

\firstsection % if your document starts with a section,
\section{Introduction}
Rotation impacts the evolution of massive stars, affecting their evolution, chemical yields,
%Rotation is a key parameter in the evolution of massive stars affecting their evolution, chemical yields, 
budget of ionizing photons and final fate as supernovae and long gamma-ray bursts \citep[e.g.][]{langer2012}. 
For massive stars the {\em initial} distribution of spin rates is especially interesting because so little is known about how these stars form \citep[e.g.,][]{zinnecker2007}.  Potentially, it can tell us more about the formation process.
%Rotation is also a key parameter in the evolution of massive stars affecting their evolution, chemical yields, budget of ionizing photons
%and their final fate as supernovae and long gamma-ray bursts. 

The 30\,Dor starburst region in the Large Magellanic Cloud contains the richest sample of massive stars in the Local 
Group and is the best possible laboratory to investigate aspects of the formation and evolution of massive stars, and to establish statistically meaningful
distributions of their physical properties. 
%
%
%In this paper, we report on the measured projected rotational velocity, $v_{e}\sin i$, for more than
%300 O-type objects observed as part of the VLT-FLAMES Tarantula Survey \citep[VFTS,][]{evans}.
%
%Up to now, studies of the spin properties of massive stars have focussed on the presumed-single star population
%with only some reporting $v_e\sin\,i$ measurements of spectroscopic binaries \citep[e.g.][]{penny,penny2009}.
%
Here we present an analysis of the $v_e\sin i$ properties of the O-type objects, i.e. spectroscopic single stars and  binary systems observed in the context of
the VLT-FLAMES Tarantula Survey \citep[VFTS;][]{evans,dekoter1}. VFTS is a multi-epoch spectroscopic campaign
targeting over 300 O-type objects -- singles and binaries. %and hundreds of cooler stars in the 30\,Dor. 
The spectral classification and radial velocity (RV) measurements, relevant for the study at hand, are presented in \citet{walborn2014}
and  \citet[][]{sana}. Here we will report on the projected rotational properties, $v_e\sin i$, of the presumed-single O-type stars \citep[][]{ramirezagudelo,ramirezagudelorhodes} 
and primary stars, i.e. the brightest component of binary systems composed of at least one O-type star (Ram\'irez-Agudelo et al. in prep.).

\section{Sample and Method}

The VFTS project and the data have been described in \citet{evans}.
In short the total Medusa sample contains $\sim$330 O-type objects. \citet[][]{sana}, 
from multi-epoch radial velocity (RV) measurements have identified  $\sim$210 O-type stars that show no significant RV
variations  ($\rm{\Delta RV}$) and are presumably single ($\rm{\Delta RV}\, \leq\, 20\, \rm{km s^{-1}}$) and $\sim$110 objects (the rest of the sample)
with $\rm{\Delta RV}\,>$ 20~$\rm{km s^{-1}}$ that are considered spectroscopic binaries.\newline 
%In the binary sample we have 85 single-lined spectroscopic binary (SB1) systems and 31 are double-lined spectroscopic binaries (SB2). 
For the single sample we use Fourier transform \citep{gray,simon} and line profile fitting methods \citep{simon2014}  to measure projected rotational velocities.
A discussion of the methods used and the accuracy that can be achieved   can be found in \citet{ramirezagudelo}. 
To obtain the projected rotational velocities of the binary sample we 
calibrate full width at half maximum or FWHM measurements versus $v_e\sin i$ for specific spectral lines, using the FWHM measurements of the O-type stars from \citet{sana}
and the $v_e\sin i$ measurements of the single O-type star presented in \citet{ramirezagudelo}. 
Details of the method and its accuracy  are found in (Ram\'irez-Agudelo et al. in prep.).

\section{Results}

%The distribution of projected rotational velocities of our single sample shows a two-component structure: a low-velocity peak at around 100 $\rm{km s^{-1}}$ and a high velocity-tail starting at about 300 kms$^{-1}$ and %extending up to $\sim$600 $\rm{km s^{-1}}$ (see Fig.~\ref{fig:dist_single_primaries}). In the low-velocity peak, these stars thus rotate at less than 20\% of their break-up velocity. For the bulk of the sample, mass loss in a %stellar wind and/or envelope expansion is not efficient enough to significantly spin down these stars \citep{vink2010,brotta}. 
%%If massive-star formation results in fast rotating stars at birth, as one may think from angular momentum conservation considerations, 
%%most massive stars have to spin down quickly by some other mechanism to reproduce the observed distribution.
%As for the rapidly rotating stars, the presence of a well populated high-velocity tail (10\% of the single sample with $v_{e}\sin i$ $>$ 300 $\rm{km s^{-1}}$) is compatible with predictions from binary evolution. \citet{selma} show that such a tail in the $v_{e}\sin i$ distribution arises naturally from mass transfer and mergers where rapid rotators result from spin-up through mass transfer and mergers. 
%1-This has important implication for the evolutionary origin of the progenitors of long gamma-ray bursts.  
%2-These progenitor systems may be dominated by, or be exclusively due to, post-interacting binaries or mergers.

\begin{figure}
\centering
\includegraphics[scale=0.35]{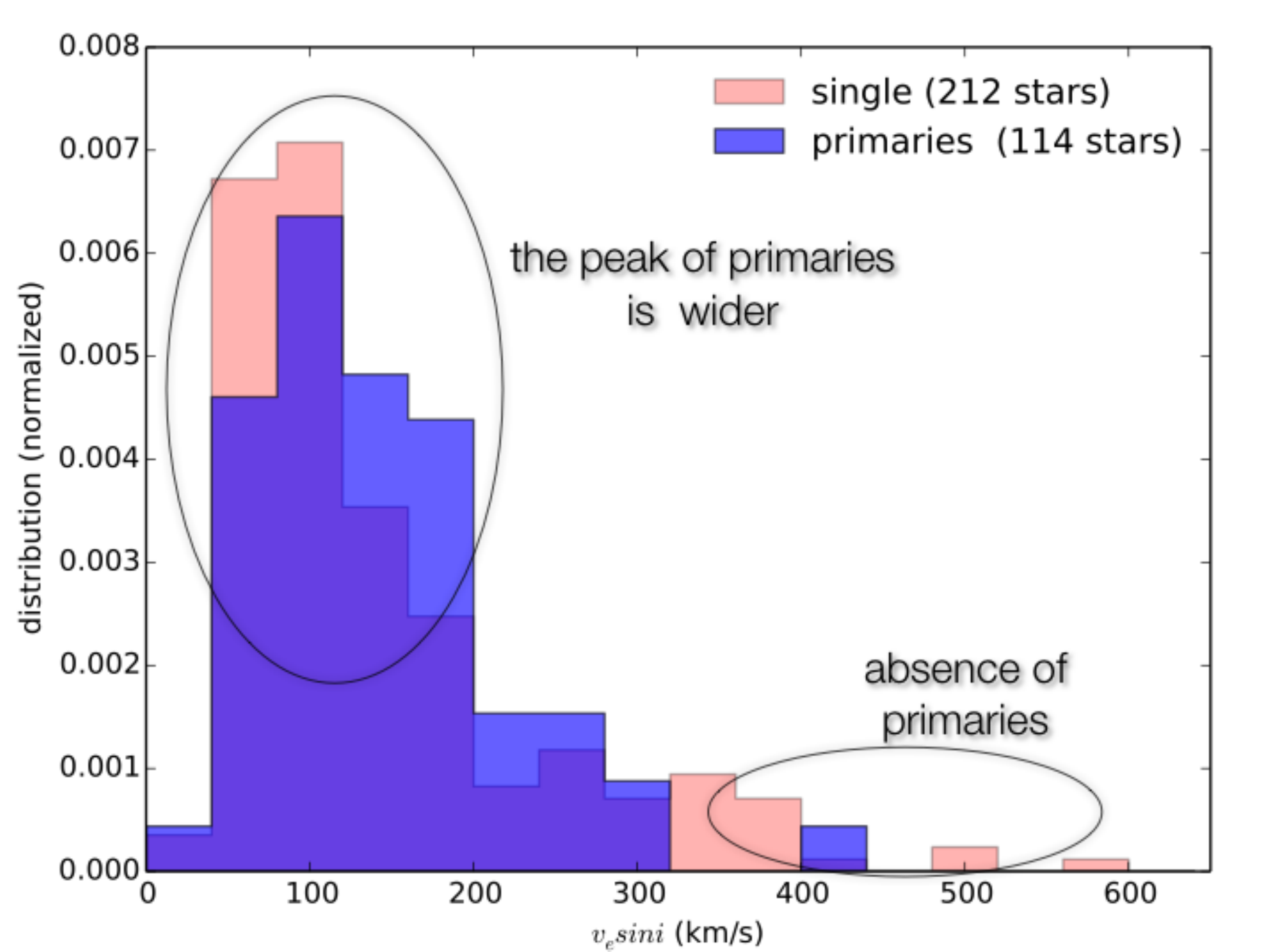}
\caption{Distribution of the projected rotational velocities, $v_e\sin i$, of the O-type presumably single and primary samples.}	
\label{fig:dist_single_primaries}
\end{figure}

Figure~\ref{fig:dist_single_primaries} shows the $v_e\sin i$ distributions of presumed-single stars and primaries. % by means of the calibration that we have adopted
Qualitatively, both distributions display a peak at around $\sim100\,\rm{km s^{-1}}$ though we also note some differences. 
First, the main peak of the primary distribution is wider than that of the presumed-single sample. 
%and overpopulates the distribution in the region between 100-300 $\rm{km s^{-1}}$ by 18\%. 
Second, at $v_e\sin i\, >$ 300\, $\rm{km s^{-1}}$ there is a deficiency of rapidly rotating
primaries with respect to the presumed-single stars. While for the presumed-single sample 22 stars out of 212  exhibit % \cancel{by almost a factor of 4}
projected rotational velocities larger than 300\, $\rm{km s^{-1}}$ (corresponding to 10$\pm$2\,\% of that sample), in the primaries we only have 3 stars out of 114 (3$\pm$1\,\%).
In Section~\ref{sec:discussion} we argue that the high-velocity tail in the presumed-single star distribution is compatible
with post-interaction binary evolution.

%\textbf{The finding of rapidly rotating stars in the single sample, i.e. high-velocity tail, is compatible withpredictions of binary evolution (see Sect.~\ref{sec:discussion})}. 
%In the discussion section we will discuss possible explanations for the differences.

%%%%%%%%%%%%%%%%%%%%%%%%%%%%%%%%%%%%%%%%%%%%%%%%%%%%%%%%%%%%%%%
%\begin{figure}
%\centering
%\includegraphics[scale=0.32]{dist_single_calibration}
%%\includegraphics[scale=0.32]{dist_single_binaries_secondaries_new}
%\caption{Distributions of the projected rotational velocities of the O-type single, primaries and secondaries.
%%($\rm{\Delta RV}\, \leq$ 20~\kms\,) 
%}	
%\label{fig:dist_single_secondaries}
%\end{figure}
%%%%%%%%%%%%%%%%%%%%%%%%%%%%%%%%%%%%%%%%%%%%%%%%%%%%%%%%%%%%%%%

%%%%%%%%%%%%%%%%%%%%%%%%%%%%%%%%%%%%%%%%%%%%%%%%%%%%%%%%%%%%%%%
%%%%%%%%%%%%%%%%%%%%%%%%%%%%%%%%%%%%%%%%%%%%%%%%%%%%%%%%%%%%%%%

\begin{figure}
\centering
\includegraphics[scale=0.28]{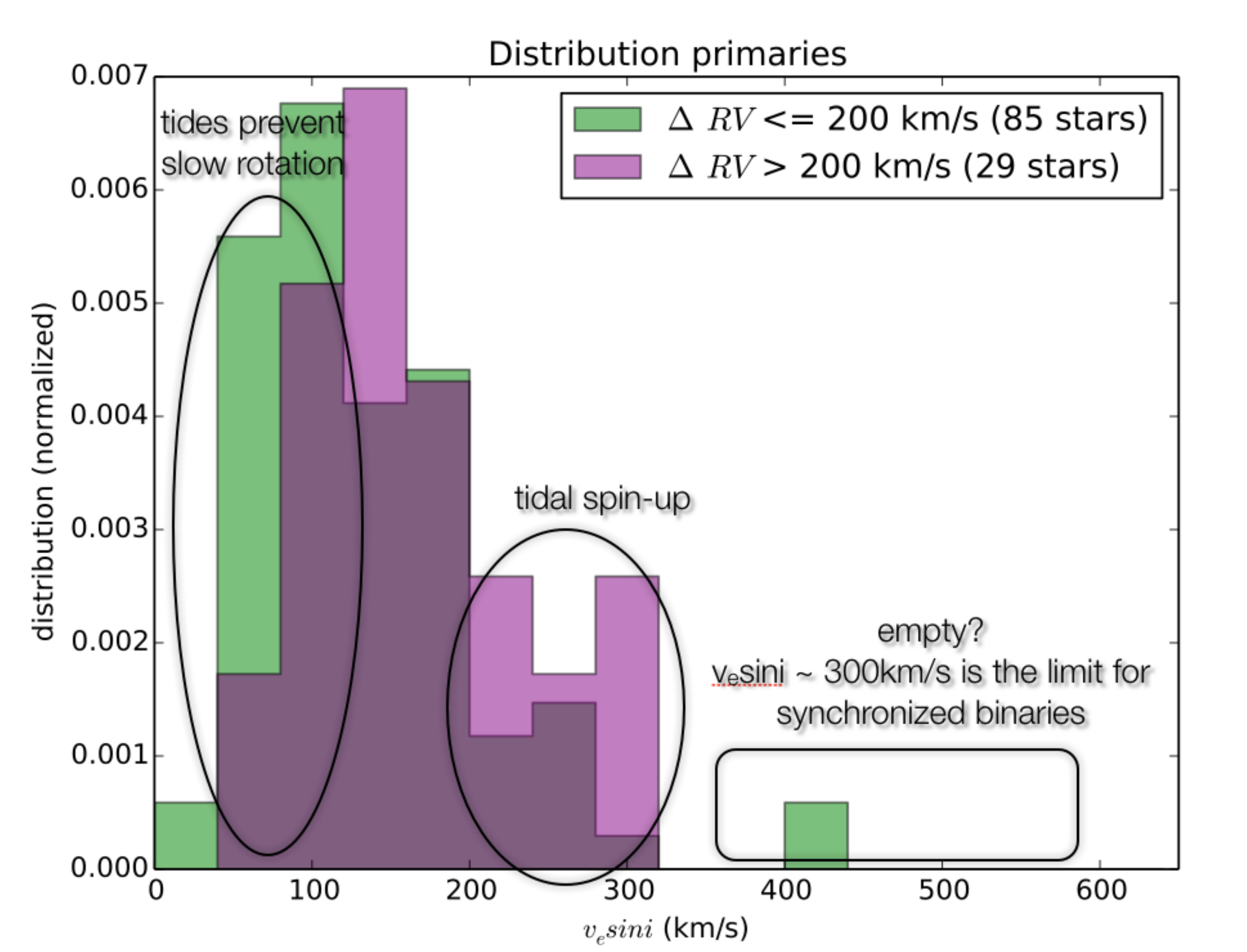}
\caption{
Distribution of the projected rotational velocities, $v_e\sin i$, of the O-type 
primaries  with  $\rm{\Delta RV}\leq \rm{\Delta RV_{limit}}$ (low-$\rm{\Delta RV}$) and   $\rm{\Delta RV}> \rm{\Delta RV_{limit}}$
(high-$\rm{\Delta RV}$).
}	
\label{fig:dist_primaries_deltarv}
\end{figure}

The size of the primary sample allows us to explore $v_e\sin i$ distributions for subpopulations.
By selecting a radial velocity amplitude limit ($\rm{\Delta RV_{limit}}$) of 200\,$\rm{km s^{-1}}$ we divide the primary sample into two subsamples,
i.e. stars that display $\rm{\Delta RV}\leq \rm{\Delta RV_{limit}}$ (henceforth low-$\rm{\Delta RV}$) and  stars with $\rm{\Delta RV}> \rm{\Delta RV_{limit}}$ (high-$\rm{\Delta RV}$).
The latter are systems where we expect tidal synchronization to become 
important during the main sequence phase; given our sample properties they roughly correspond to 
binaries with an orbital period of less than 10\,days (see Ram\'irez-Agudelo et al. in prep.).
Most of the systems in the low-$\Delta$RV subsample 
are wider systems that
will not suffer from tides with the potential exception of those that are seen at
a low inclination angle. Figure~\ref{fig:dist_primaries_deltarv} plots the $v_e\sin i$ distributions of the low-$\rm{\Delta RV}$ (85 stars) and high-$\rm{\Delta RV}$ (29 stars)
subsamples. At $v_e\sin i$ $\leq$ 200\, $\rm{km s^{-1}}$ the distribution of high-$\rm{\Delta RV}$ sources,
compared to the low-$\rm{\Delta RV}$ sources, appears shifted by one bin to higher rotational velocities. 
%\textit{As a result, there are 13$\pm$11\,\% fewer high-$\rm{\Delta RV}$ sources up to $v_e\sin i$ = 200\, $\rm{km s^{-1}}$ compared to low-$\rm{\Delta RV}$ sources.
%Second, the high-$\rm{\Delta RV}$ distribution overpopulates the distribution in the region between 200 and 300\, $\rm{km s^{-1}}$
% by 16$\pm$11\,\% with respect to the low-$\rm{\Delta RV}$ distribution. %(27\% of the former compared to 11\%, respectively ).
%Third, in the high-velocity tail (i.e. $v_e\sin i$ $>$ 300 $\rm{km s^{-1}}$) the low-$\rm{\Delta RV}$ distribution has three primaries (4$\pm$2\,\% of that sample), while  
%there is no star in the high-$\rm{\Delta RV}$ distribution}.
The weighted mean of the samples (119 and 190\,$\rm{km s^{-1}}$ 
for the low- and high-$\rm{\Delta RV}$ samples respectively) are significantly different from one another, confirming an average faster rotation
for the high-$\rm{\Delta RV}$ (short periods) systems.
These differences may be related to tidal effects, an hypothesis that we will explore further in the next subsection.

% \textbf{This difference 
%%and that the low-velocity peak of the primary distribution is wider that than of the single sample
%may be related due to the effects of tides. In next subsection we explore further this scenario.}

%%%%%%%%%%%%%%%%%%%%%%%%%%%%%%%%%%%%%%%%%%%%%%%%%%%%%%%%%%%%%%%
%%%%%%%%%%%%%%%%%%%%%%%%%%%%%%%%%%%%%%%%%%%%%%%%%%%%%%%%%%%%%%%

\subsection{Tidal interaction} \label{subsec:tides}
\begin{figure}
\centering
\includegraphics[scale=0.48]{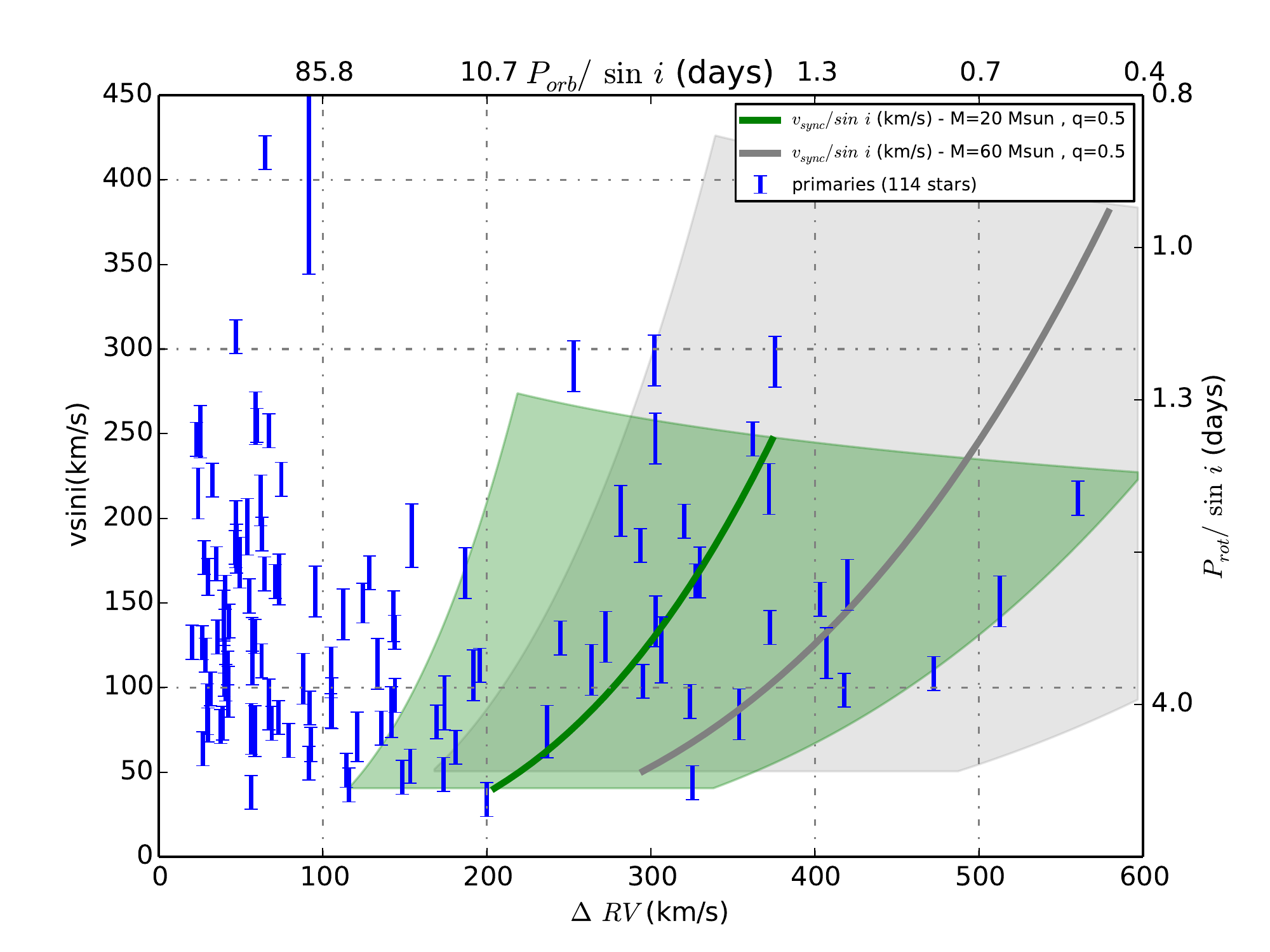}
\caption{Projected rotational velocity vs. radial velocity amplitude for the primary sample (114 stars). The green and gray 
regions show where a 20\,$M_{\odot}$ and 60\,$M_{\odot}$ primary is expected to become synchronized with its companion before the
primary leaves the main sequence, computed for mass ratios $q$ ranging between 0.25 and 1 and 
periods $P_{\rm orb}$ from 0.25 to 10 days. The thick green and gray lines are for $q = 0.5$.
The top axis displays the orbital period $P_{\rm orb}/ \sin i$ associated
with the radial velocity amplitude given on the horizontal axis for a 20\,$M_{\odot}$ primary.  Similarly, $P_{\rm rot}/ \sin i$ displayed on the right vertical axis corresponds to the projected 
rotational velocity given on the left. % (Color version available online). 
}
\label{fig:vsini_deltarv}
\end{figure}

To investigate the effect of tidal locking in our sample we make use of a diagram showing the relation between projected rotational velocity 
and amplitude of the RV variations (see Fig.~\ref{fig:vsini_deltarv}).
 The timescale of tidal synchronization is a function of the primary mass $M_{1}$, the mass ratio of the primary and secondary $q = M_{2}/M_{1}$, and
the orbital period $P_{\rm orb}$.  For our sample we lack information about the latter and therefore we use the radial velocity amplitude
$\Delta$RV as a proxy of the period. Figure~\ref{fig:vsini_deltarv} gives the relation between $v_e\sin i$ and $\Delta$RV for the
sample of primaries. The green dashed region shows the range of parameter space for which tidal synchronization of
a 20\,$M_{\odot}$ primary -- a typical mass for our sample -- occurs within the main sequence life for $q$ ranging from
0.25 (left-side boundary) to 1 (right-side boundary) and $P_{\rm orb}$ ranging from 10 d (bottom boundary) to
0.25 d (upper boundary), or sooner when the primary fills its Roche lobe prior to reaching the end of the main sequence
\citep{selma2009b,selma}. The thick green line is for $q = 0.5$. The top axis of the figure displays the orbital period $P_{\rm orb}/ \sin i$ 
associated with the radial velocity amplitude given on the horizontal axis, assuming $\Delta RV$ is twice the actual 
semi-amplitude of the radial velocity curve.  Similarly, the $P_{\rm rot}/ \sin i$ displayed on the right vertical axis 
corresponds to the projected  rotational velocity given on the left. 

Tides are less effective for smaller mass ratios and longer periods, therefore the timescale of tidal synchronization is
longest for systems in the lower left corner of the green zone. In the upper right corner synchronization proceeds the
fastest. For instance, for a 0.5\,d system with $q = 0.75$ synchronization by turbulent viscosity and radiative dissipation 
occurs within about 1 percent of the main sequence lifetime \citep{selma2009b}.
The gray region shows the $M_{1}$ dependence by also displaying the zone as defined here 
for a 60\,$M_{\odot}$ primary. It is stretched out to both higher $v_e\sin i$ $(\propto R)$, and 
$\Delta RV$ $(\propto M_{1}^{1/3})$, where $R$ is the stellar radius. 
The green zone that according to the above argument may be expected to contain systems that are synchronized or tend
toward synchronization is populated by 40 primaries.
The 29 sources that constitute the pink ($\rm{\Delta RV} > 200 \,\rm{km s^{-1}}$) distribution in 
Fig.~\ref{fig:dist_primaries_deltarv} are almost exclusively
found in the green region, more specifically in the part of that region for which synchronization by tides is the most relevant.

%%%%%%%%%%%%%%%%%%%%%%%%%%%%%%%%%%%%%%%%%%%%%%%%%%%%%%%%%%%%%%%
%%%%%%%%%%%%%%%%%%%%%%%%%%%%%%%%%%%%%%%%%%%%%%%%%%%%%%%%%%%%%%%

\section{Implications}\label{sec:discussion}
The most distinctive feature of the $v_e\sin i$ distribution of the O-type presumed-single stars and primaries
in 30 Dor is its  low-velocity peak at around 100\,$\rm{km s^{-1}}$. 
For the bulk of the samples, mass loss in a stellar wind and/or envelope expansion is not efficient enough to significantly spin down these stars.
Therefore the peak is likely to be the outcome of the formation process (see discussion in Ram\'irez-Agudelo et al. 2013 and in prep.).
The presence of a high-velocity tail in the presumed-single sample, and the absence of such a tail in the primary sample, is compatible with predictions of binary interaction. 
Rapid rotators result from spin-up through mass transfer and mergers \citep{selma}, that mostly appear as, or have become, single objects \citep{selma2014a}.  Such a nature of the high-velocity tail has 
important implications for the evolutionary origin of the progenitors of long gamma-ray bursts, reducing the likelihood that long-GRBs may also result from single stars that are born spinning rapidly. 
Finally, if post-interaction systems have been removed from the binary sample for the above outlined reason it is dominated by pre-interaction systems. The short period systems among these may suffer from tidal
synchronization effects that may qualitatively explain the differences in the low-velocity peak structure of presumed-single stars and of primaries.  

\bibliographystyle{iau307}
%\bibliography{MyBiblio}

%\newpage
\begin{discussion}

\discuss{Phil Massey}{Conti \& Ebbets (1977) showed that the dwarf O stars had both a low and a high $v_e\sin i$ peak but that the O supergiants have only a low peak. How can
post interacting O stars (merged or not) be luminosity class V with modest $\rm{M_v}$? Would not you expect them to have high luminosities?}

%\discuss{Oscar}{That a given O-type star is of luminosity class (LC) I does not necessarily mean it's a supergiant, it could still be in the hydrogen burning phase (i.e. main sequence star). 
%In that way the presence of O stars with LCI and high $v_e\sin i$ is possible. This is what we find and discuss in  Sect. 4 of \citet[][]{ramirezagudelo}.} 

\discuss{Oscar}{In  \citet[][]{ramirezagudelo} subdivide the contribution of dwarfs, giants and supergiants to the $v_e \sin i$ distribution.  We find that the tail is dominated by the
dwarfs and that supergiants are almost missing. The reason is that the break-up velocity of supergiants is considerably lower, about 300\,$\rm{km s^{-1}}$, than that of dwarfs ($\sim 600\,\rm{km s^{-1}}$), simply 
because they are bigger. The secondary O\,V stars that receive the mass may remain dwarfs, the interaction itself does not turn them into supergiants.  Later they may evolve towards
supergiants -- but then their $v_e \sin i$ will decrease because of envelope expansion. } 

\discuss{Andre Maeder}{You have shown that you are calibrating your $v_e\sin i$ on the basis of the change of the line profiles. However, as shown long ago by Collins, the equivalent
width of the lines may also change since the stellar flux is coming from regions with a variety of $\rm{g_{eff}}$ and $\rm{T_{eff}}$. If this is not accounted for, a large underestimate of the 
$v_e\sin i$ may result. Could you please comment on what you are doing.}

%\discuss{Oscar}{Indeed, it may be the case. To determine $v_e\sin i$, for the single sample,  we explore the reliability of the diagnostic lines. In this way, 
%we have predicted an underestimation by 20\% of HeII4541 compared to the HeI lines for high spin rates \citep[$v_e\sin i >$ 250\kms,][see Sect. 3]{ramirezagudelo}.
%Similarly, gravity darkening may also cause this. In early B-type stars \citet{towsend} found that gravity darkening may underestimate the true $v_e\sin i$ up to 20\% but a similar studies has
%not been undertaken for O-type stars.}

\discuss{Oscar}{The gravity darkening effect that you refer to has been studied for B stars by \citet{towsend} and becomes important for stars that spin faster than about 95\% of critical.
In our sample of presumably-single O stars only six sources out of the $\sim$210 spin above 60\% of critical, and one source out of 114 binaries.  Because of the very limited effect on the
sample as a whole we did not correct for the darkening effect.}

\end{discussion}

\end{document}